\documentclass[onecolumn,manuscript,preprint,aps,pra]{revtex4}

\usepackage{bm,amsmath,amssymb,stmaryrd} \usepackage{graphicx}
\usepackage{amssymb}
\usepackage{epsfig}
\usepackage[english]{babel}
\usepackage{latexsym}
\usepackage{graphics}
\usepackage{subfigure}
\usepackage{epsfig}
\usepackage{graphicx}
\usepackage{dcolumn}
\usepackage{amsmath}

\begin{document}
\title{Angle-Resolved Spectra of the Direct Above-Threshold Ionization of Diatomic Molecule in IR+XUV laser fields\thanks{Project supported by the National Natural Science Foundation of China (Grant No.~11334009, 11425414, 11474348 and 11774411).}}

\author{Shang Shi$^{1,3}$, \ Facheng Jin$^{2}$,  \ and \ Bingbing Wang$^{1}$\thanks{Corresponding author. E-mail:~wbb@aphy.iphy.ac.cn}\\
$^{1}${\emph{Laboratory of Optical Physics, Beijing National Laboratory for Condensed Matter Physics},}\\
{\emph{Institute of Physics, Chinese Academy of Sciences, Beijing 100190, China}}\\  
$^{2}${\emph{Faculty of Science, Xi'an Aeronautical University, Xi'an 710077, China}}\\ 
$^{3}${\emph{University of Chinese Academy of Sciences, Beijing 100049, China}}}   

\date{\today}

\begin{abstract}
    The direct above-threshold ionization (ATI) of diatomic molecules in linearly-polarized infrared and extreme ultraviolet (IR+XUV) laser fields  is investigated by the frequency-domain theory based on the nonperturbative quantum electrodynamics. The destructive interference fringes on the angle-resolved ATI spectra, which are closely related to the molecular structure, can be well fitted by a simple predictive formula for any alignment of the molecular axis. By comparing the direct ATI spectra for monochromatic and two-color laser fields, we found that the XUV laser field can both raise the ionization probability and the kinetic energy of the ionized electron, while the IR laser field can broaden the energy distribution of the ionized electron.  Our results demonstrate that, by using IR+XUV two-color laser fields, the angle-resolved spectra of the direct ATI can image the structural information of molecules without considering the recollision process of the ionized electron.
\end{abstract}

\par\noindent

\maketitle

\section{Introduction}

In the past decade, two important processes of nonlinear dynamics in intense laser field$\raisebox{0.3mm}{---}$high-order harmonic generation (HHG) in which the ionized electrons recombine with the parent ion \cite{chu,gao2000} and high-order above-threshold ionization (HATI) in which the ionized electrons elastically collide with the parent ion \cite{DBM2018,Dino2018,Wang2007}$\raisebox{0.3mm}{---}$have been widely used in the study of the molecular structure, since the molecular structure is of great significance in the fields of physics, biology and chemistry. For homonuclear diatomic molecules, the results of the theories \cite{chun2009,Guo2009,becker1,Hetzheim2007} and experiments \cite{kang2010,becker2} of HATI process indicate that the HATI spectrum carries the important information of molecules, such as internuclear separation, the symmetry of the highest occupied molecular orbital as well as the alignment of molecular axis. However, the interference structure in the HATI spectra of heteronuclear diatomic and polyatomic molecules \cite{Wang2010,ingo2015,Odzak2014} are more complicated than that of homonuclear diatoms. One needs to consider the influence of the charge distribution of molecular ions on the interference fringes \cite{Wang2010}. Recently, the studies on the HHG process of molecules also demonstrate theoretically \cite{Odzak2014,toru2008,odzak2009,Augstein2011,Su2018} and experimentally \cite{Torres2010} that the HHG spectra can be used to expose the structure and dynamics of molecules.

Unlike the two-step processes of HHG and HATI, the direct ATI is a one-step process where the electrons reach the detector directly after the ionization \cite{EBERLY1991,BECKER2008}. Therefore, the ionization spectra of the direct ATI can directly reflect the initial information of the molecular properties. The results obtained by He \textit{et al.} \cite{he2015} who used extreme ultraviolet (XUV) pulses to study the direct ATI of ${\rm H}_2^+$ demonstrate that the initial probability distribution of molecules can be imaged from the angular distribution of the directly ionized electron. However, most studies do not apply the direct ATI process to explore the molecular structure, since the structural information of a molecule is mainly embodied in the high-energy region of the ionization spectra where only the re-colliding electron may dominate \cite{becker1,Hetzheim2007,kang2010,becker2} by applying a commonly used infrared (IR) laser field.

In order to gain the angle-resolved spectra of direct ATI which can fully exhibit the structural information of molecules, we consider a combined IR and XUV laser fields to investigate the direct ATI process of a diatomic molecule. The idea of applying IR+XUV laser fields to the study of ultrafast nonlinear dynamics has already existed, see Ref. \cite{kui2017,Richard2008,Liu2015,Zhang2013}. In our results, the destructive interference fringes related to the molecular structure are clearly displayed in the angle-resolved spectra of the direct ATI. By comparing the direct ATI spectra of hydrogen molecule in two-color laser field with that in a monochromatic laser field, we found the different roles of IR and XUV laser fields in the ionization of an electron. The study on oxygen molecules further manifests the universality of the method which utilizes the angle-resolved spectra of the direct ATI in IR+XUV two-color laser fields to infer the molecular structure.

\section{Theoretical method}

By using the frequency-domain theory, we deal with the direct ATI process of a diatomic molecule in two-color linearly polarized laser fields with laser frequencies $\omega_1$ for low-frequency field and $\omega_2$ for high-frequency field. The Hamiltonian of the molecule-radiation system is (atomic units are used throughout unless otherwise stated)
\begin{equation}\label{eq21}
H=H_0+U(\textbf{r})+V,
\end{equation}
where
\begin{equation}\label{eq22}
 H_0=\frac{(-i\nabla)^2}{2}+\omega_1N_{a1}+\omega_2N_{a2},
\end{equation}
is the free-electron and free-photon energy operator and $N_{aj}=(a^\dag_ja_j+a_ja^\dag_j)/2$ ~($j=1,2$) is the photon number operator  with $a_j(a^\dag_j)$ the annihilation (creation) operator of the laser photon mode. $U(\textbf{r})$ is the two-center binding potential of a diatomic
molecule. $V$ is the electron-photon interaction which is expressed as
\begin{equation}\label{eq23}
 V=-[(-i\nabla)\cdot \textbf{A}_1(-\textbf{k}_1\cdot \textbf{r})+(-i\nabla)\cdot \textbf{A}_2(-\textbf{k}_2\cdot \textbf{r})]+\frac{1}{2}[\textbf{A}_1(-\textbf{k}_1\cdot \textbf{r})+\textbf{A}_2(-\textbf{k}_2\cdot \textbf{r})]^2,
\end{equation}
where $\textbf{A}_j(-\textbf{k}_j\cdot \textbf{r})=(2V_{\gamma_j}\omega_j)^{-1/2}({\hat{\bm{\epsilon}}}_j e^{i\textbf{k}_j\cdot \textbf{r}}a_j+c.c.)$  is the vector potential with $\textbf{k}_j$ the wave vector,~${\hat{\bm{\epsilon}}}_j=\hat{z}$ the polarization vector and $V_{\gamma_j}$ the normalization volume of the laser field with frequency $\omega_j$.

The initial state of the system is $|\psi_i\rangle=|\Phi_i(\textbf{r}),l_1,l_2\rangle=\Phi_i(\textbf{r})\otimes |l_1 \rangle \otimes |l_2 \rangle$, which is the eigenstate of the Hamiltonian $H_0+U(\textbf{r})$ with energy $E_i=-E_B+(l_1+\frac{1}{2})\omega_1+(l_2+\frac{1}{2})\omega_2$. Here, $\Phi_i(\textbf{r})$ is the ground-state wave function of the molecule with binding energy $E_B$, and $|l_j \rangle$ is the Fock state of the laser mode with photon number $l_j$. On the other hand, the final state $|\psi_f\rangle=|\psi_{\textbf{p}_fn_1n_2}\rangle$ is the quantized-field Volkov state in two-color laser fields~\cite{Guo1992}
\begin{equation}\label{eq2}
\begin{array}{l}
 {|\psi_{\textbf{p}_f n_1 n_2}(\textbf{r})\rangle} = V_e^{ - 1/2}\exp [i(\textbf{p}_f+u_{p_1}\textbf{k}_1+u_{p_2}\textbf{k}_2)\cdot\textbf{r}]\\       ~~~~~~\times\sum\limits_{\substack{{q_1} =  - {n_1}\\{q_2} =  - {n_2}}}^\infty \exp \{-i[q_1(\textbf{k}_1\cdot\textbf{r}+\phi_1)+q_2(\textbf{k}_2\cdot\textbf{r}+\phi_2)\} \\
~~~~~~\times\mathcal{J}_{q_1 q_2}(\zeta_f)^{*}|n_1+q_1,n_2+q_2\rangle, \\
 \end{array}
\end{equation}
which is the eigenstate of the Hamiltonian $H_0+V$ with eigenvalue $E_{\textbf{p}_fn_1n_2}=\frac{\textbf{p}_f^2}{2}+(n_1+\frac{1}{2}+u_{p_1})\omega_1+(n_2+\frac{1}{2}+u_{p_2})\omega_2$, where $u_{p_j}$ is the ponderomotive energy in unit of one photon energy of the laser field. In Eq. (\ref{eq2}), $q_1$ and $q_2$  represent the number of photons absorbed by the electron from IR and XUV laser fields, respectively. $\textbf{p}_f$ is the final momentum of the ionized electron. $\phi_1$ and $\phi_2$ are the initial phases of each laser field and are taken as zero for simplicity in this paper. The term $\mathcal{J}_{q_1 q_2}(\zeta_f)$ is the generalized Bessel function, which is expressed as
\begin{equation}\label{eq3}
\begin{array}{l}
\mathcal{J}_{q_1 q_2}(\zeta_f)=\sum\limits_{q_3 q_4 q_5 q_6} J_{-q_1+2q_3+q_5+q_6}(\zeta_1)J_{-q_2+2q_4+q_5-q_6}(\zeta_2)J_{-q_3}(\zeta_3)\\
~~~~~~~~~~~\times J_{-q_4}(\zeta_4)J_{-q_5}(\zeta_5)J_{-q_6}(\zeta_6),
 \end{array}
\end{equation}
where the  arguments of the generalized bessel function are as follows
\begin{equation}\label{eq4}
\begin{array}{l}
\zeta_1=2\sqrt{\displaystyle\frac{u_{p_1}}{\omega_1}}p_f\cos\theta_f,~\zeta_2=2\sqrt{\displaystyle\frac{u_{p_2}}{\omega_2}}p_f\cos\theta_f,\\
\zeta_3=\frac{1}{2}u_{p_1},~~~~~~~~~~~~~~~\zeta_4=\frac{1}{2}u_{p_2},\\
\zeta_5=2\displaystyle\frac{\sqrt{u_{p_1}u_{p_2}\omega_1\omega_2}}{\omega_1+\omega_2},~~\zeta_6=2\displaystyle\frac{\sqrt{u_{p_1}u_{p_2}\omega_1\omega_2}}{\omega_1-\omega_2}.
 \end{array}
\end{equation}
Here, we define $\theta_f$ as the angle of the momentum direction of the final electron with respect to the polarization direction of the laser fields.

The transition matrix element of the direct ATI process from the initial state $|\psi_i\rangle$ to the final state $|\psi_f\rangle$ is written as \cite{Wang2007}
\begin{equation}\label{eq1}
T_{fi} =\langle \psi_f|V|\psi_i\rangle.
\end{equation}
By inserting the expressions of the initial state and final state into Eq.~(\ref{eq1}), the transition matrix element becomes
\begin{equation}\label{eq5}
T_{fi} =V_e^{-1/2}[(u_{p_1}-q_1)\omega_1+(u_{p_2}-q_2)\omega_2]\mathcal{J}_{q_1 q_2}(\zeta_f)\Phi_i(\textbf{p}_f),
\end{equation}
where $\Phi_i(\textbf{p}_f)$ is the fourier transform of the ground-state wave function of the diatomic molecule. For hydrogen molecule, the ground-state wave function was obtained by a
linear combination of two atomic wave functions with Gaussian forms, hence we have\\
\begin{equation}\label{eq6}
\begin{array}{l}
\Phi_i(\textbf{p}_f)=2\cos(\displaystyle\frac{p_fR_0\cos\theta}{2})\pi^{-3/4}e^{-\textbf{p}_f^2/2}\frac{1}{\sqrt{2(1+e^{-R^2_0/4})}}.
 \end{array}
\end{equation}
Here, $R_0$ is  the nuclear spacing of the molecule, $\theta$ is the angle between the momentum direction of the ionized electron and the molecular axis, and $\cos\theta$ is expressed as $\cos\theta=\sin\theta_f\cos\varphi_f\sin\theta_m\cos\varphi_m+\sin\theta_f\sin\varphi_f\sin\theta_m\sin\varphi_m+
\cos\theta_f\cos\theta_m$, where $\varphi_m$ and $\varphi_f$ are the azimuthal angles of the molecular axis and the final momentum of the electron, respectively. $\theta_m$ is the angle between the polarization vectors of the laser field and the molecular axis.  We take $\varphi_m=\varphi_f=0$ in the calculation of the electron energy spectra, thus $\cos\theta$ can be reduced to $\cos(\theta_f-\theta_m)$.  For oxygen molecule, the initial wave function was calculated by the GAMESS software~\cite{Wang2010}. Thus the Fourier transform of the ground-state wave function of ${\rm O}_2$ is expressed as \\
\begin{equation}\label{eq7}
\Phi_i(\textbf{p}_f)=p_f\sin(\frac{p_fR_0\cos\theta}{2})\sqrt{1-\cos^2\theta}\cos\varphi_f/(1+(\textbf{p}_f)^2)^3.
\end{equation}

\section{Results and discussion}

 We first compare the angle-resolved spectra of the direct ATI for atomic hydrogen and molecular hydrogen in linearly polarized IR+XUV laser fields with both intensities of ${\rm 3.6 \times 10^{13}~ W/cm}^2$, as shown in Figs. 1(a)  for the hydrogen atom  and  Figs. 1(b-c) for the hydrogen molecule. The photon energy of XUV laser is $\omega_2=15\omega_1$, where $\omega_1=1.55$~eV is the photon energy of IR laser. The hydrogen molecular axis is along the directions of the laser polarization. The ionization potential for atomic hydrogen and molecular hydrogen are 13.6 eV and 13.12 eV, respectively.  It can be seen that some destructive interference fringes (DIF) in the molecular spectra  do not exist in the atomic spectrum. By analyzing the transition formula of the direct ATI process (see Eq. (\ref{eq5})), we found that the  condition for the emergence of these DIF can be expressed as $\cos(p_f\cos\theta R_0/2)=0$. Thus, we obtain  the $E_f-\theta_f$ curve which can predict the position of the DIF in the angle-resolved ionization spectra, as seen in Figs. 1(b) and (c). The corresponding expression of the photoelectron energy $E_f$ is
\begin{equation}\label{eq8}
E_f=\frac{p_f^2}{2}=\frac{[(1+2n)\pi]^2}{2R_0^2\cos^2(\theta_f-\theta_m)},
\end{equation}
where $R_0$ is the internuclear separation, $n=0, \pm1, \pm2, ...$. The black dashed lines in Figs. 1(b-c) correspond to $n=0$, and the red dash-dotted line in Figs. 1(c) corresponds to $n=1$. It can be seen from  Eq. (\ref{eq8}) that the energy of the photoelectron in the DIF can get minimum value, i.e., $E_{fmin}=\pi^2/2R_0^2$, when $n=0$ with the electron emitted along the molecular axis, i.e., $\theta_f=\theta_m$. Therefore, the DIF will appear in the spectra as long as the range of the direct ATI spectra exceed the value of $E_{fmin}$. Additionally, since $E_{fmin}$ is inversely proportional to $R_0^2$, the value of $E_{fmin}$ for $R_0=6~a.u.$ decreases to one ninth of that for $R_0=2~a.u.$, hence this is the reason why the number of the DIF in Figs. 1(c) is more than that in Figs. 1(b).

\begin{center}
\includegraphics{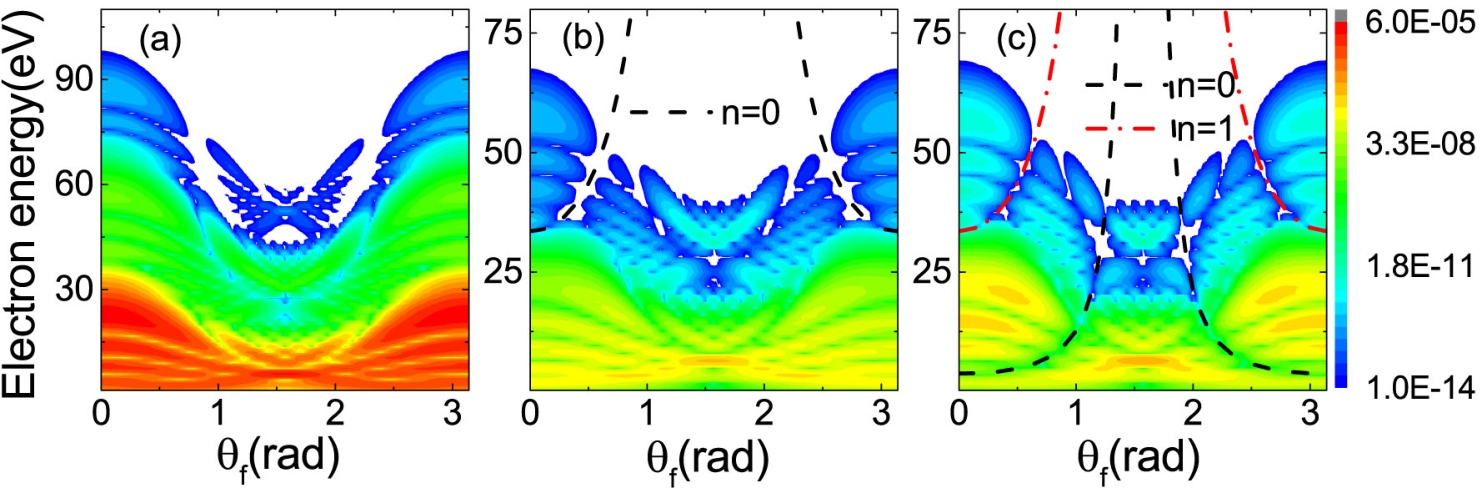}\\[5pt]  
\parbox[c]{15.0cm}{\footnotesize{\bf Fig.~1.} (color online) The angle-resolved spectra of the direct ATI for a hydrogen atom (a) and for a hydrogen molecule with the internuclear distance  $R_0=2.0~a.u.$  (b) and $R_0=6.0~a.u $  (c). The photon energy of XUV laser is $\omega_2=15\omega_1$, where $\omega_1=1.55$~eV represents the photon energy of  IR laser. Both the photon energies of IR and XUV laser remain constant for the case of ${\rm H}_2$ in this paper. The laser intensities are $I_2=I_1={\rm 3.6 \times 10^{13}~ W/cm}^2$, where $I_2$ is the intensity of XUV laser field  and $I_1$ the intensity of IR laser field. The molecular axis of ${\rm H}_2$ is along the polarization directions of two laser fields, which are fixed at z axis. In logarithmic scale.}
\end{center}

The angle-resolved  spectra of the direct ATI of ${\rm H}_2$ with different orientation angles of the hydrogen molecular axis with respect to the polarization vectors of the laser fields is shown in Fig. 2. Parameters are as in Figs. 1(b). It shows that the distribution of the angle-resolved  spectra and the DIF significantly depend on the angle between the molecular axis and the polarization direction of the laser. Compared to Ref. \cite{becker1}, where the predictive curve can  overlap well with the DIF in angle-resolved HATI spectra only when the molecular axis is perpendicular to the laser polarization, we may find that the interference stripes related to the molecular structure in the spectra presented by Fig. 2 can be perfectly reproduced by Eq. (\ref{eq8}) at any orientation angle $\theta_m$. This indicates that the angle-resolved  spectra of the direct ATI  may provide more details about the information of the molecular structure.

\begin{center}
\includegraphics{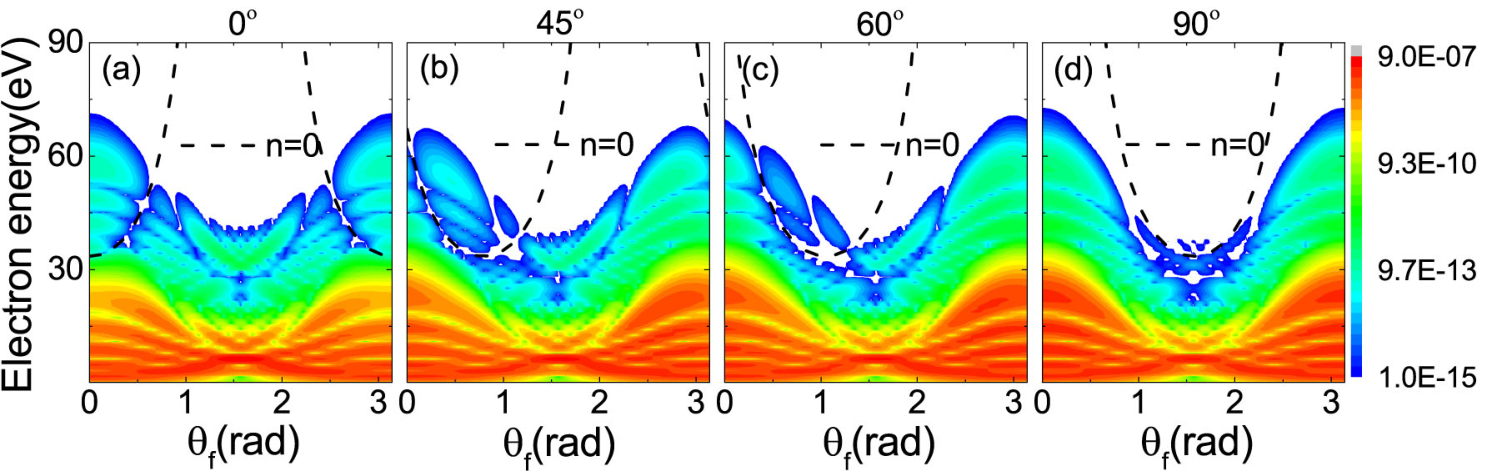}\\[5pt]  
\parbox[c]{15.0cm}{\footnotesize{\bf Fig.~2.} (color online) The angle-resolved  spectra of the direct ATI of ${\rm H}_2$ with different orientation angles  $\theta_m$=~0$^\circ$ (a), 45$^\circ$ (b), 60$^\circ$ (c) and 90$^\circ$ (d). The internuclear separation is 2~a.u. and remains unchanged in the following results of the hydrogen molecule. The intensities and the photon energy of each laser field are the same as in Fig. 1. In logarithmic scale.}
\end{center}

In order to explain the plateaulike distribution of the angle-resolved spectra of the direct ATI, we analyze the transition matrix element of the direct ATI process under the conditions of Figs. 1(b), as shown in Fig. 3. Here, $q_2$ represents the absorption number of XUV photons. Compared with $q_2=1$, the whole spectrum moves to higher energy region when $q_2=2$. By applying the energy conservation rule during the ionization process derived from the saddle-point approximation (see Appendix), we predict the classical  position of the beginning (dashed line) and the cutoff (solid line) of the ATI spectra with different values of $q_2$, which match well with the numerical calculation results.  Figure 3 demonstrates that the XUV laser field plays a crucial role in making the high-energy plateau on the direct ATI spectra, in marked contrast to the situation that the high-energy plateau is absent in  the  direct ATI spectra for IR laser case \cite{becker1,becker2}. We can also see that the DIF conspicuously appear in the angle-resolved spectra when $q_2=2$. This is because the minimal energy for the emergence of DIF is about 33.6 eV according to Eq. (\ref{eq8}), and hence the electron needs to absorb at least two XUV photons to obtain such high energy.

\begin{center}
\includegraphics{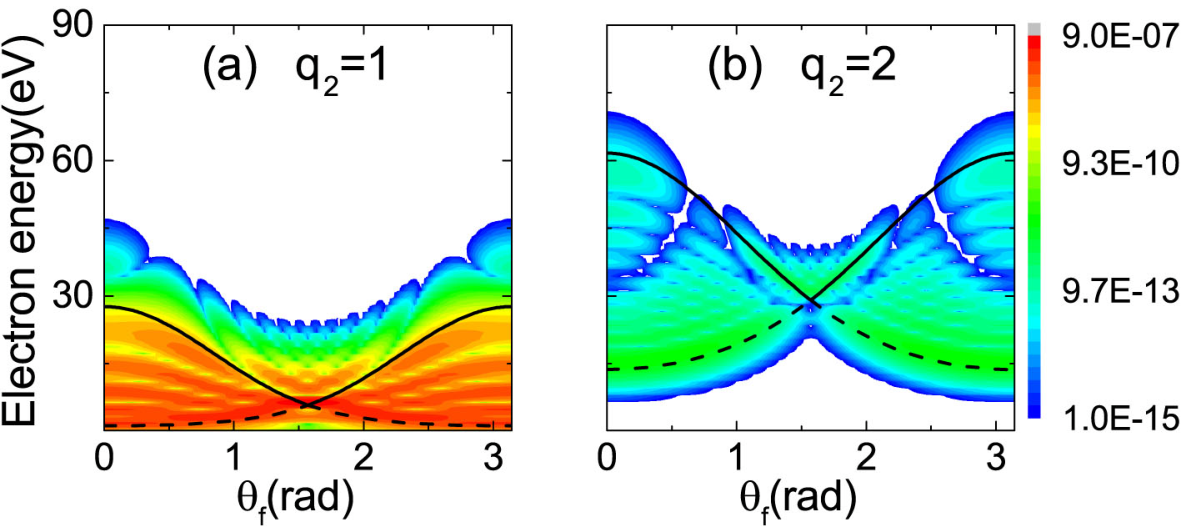}\\[5pt]  
\parbox[c]{15.0cm}{\footnotesize{\bf Fig.~3.} (color online) The angular distribution for the direct ATI  of ${\rm H}_2$ absorbing a certain number of  XUV photons, which is denoted by $q_2$. The parameters are consistent with  Fig. 1(b). The solid and dashed lines predict the cutoff and the beginning position of the ATI  spectra for different $q_2$. In logarithmic scale.}
\end{center}

\begin{center}
\includegraphics{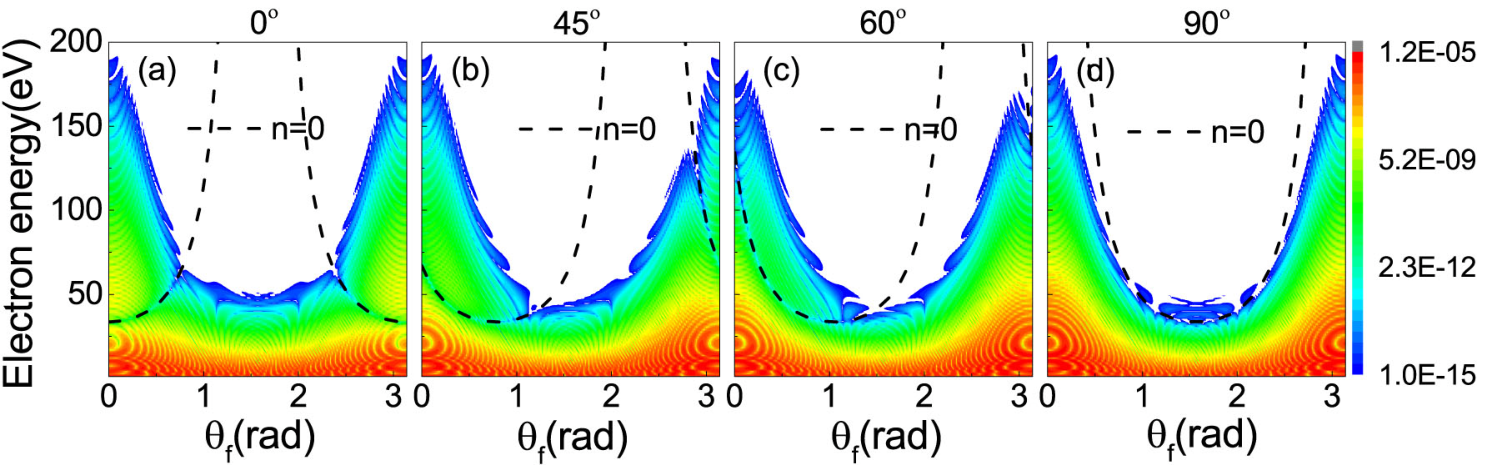}\\[5pt]  
\parbox[c]{15.0cm}{\footnotesize{\bf Fig.~4.} (color online) The angle-resolved direct ATI spectra of ${\rm H}_2$ in an IR laser filed with the intensity of ${\rm 3.6\times10^{15}~W/cm}^2$ at four orientation angles of the molecular axis. In logarithmic scale.}
\end{center}

We now consider the direct ATI of the molecular hydrogen in a monochromatic laser field. Since the term $\cos(p_f\cos\theta R_0/2)$ in the direct ATI transition formula originates from the coherent superposition of the ionization paths of the electron in the hydrogen molecule, the DIF in the spectra is actually unconcerned with whether or not the laser fields are two-color. Therefore, for monochromatic IR laser field, the DIF appear in the angular spectra of the direct ATI as the laser intensity is increased strong enough. Figure 4 presents the direct ATI spectra of monochromatic IR laser field with its intensity increased to $I_1={\rm 3.6 \times 10^{15}~ W/cm}^2$. Compared to Fig. 2 and Fig. 3, it is found that the XUV laser field can raise the  ionization probability in the ATI process. On the other hand, for the XUV laser field with $\omega_2=23.25$~eV and $I_2={\rm 3.6 \times 10^{13}~ W/cm}^2$, the distribution of the ionization spectra totally differs from that of two-color laser fields and monochromatic IR laser field. Namely, the distribution of the electron energy is no longer continuous, and there are many minimums in the spectra, as shown in Fig. 5. However, by comparing with the angle-resolved probability spectra of a hydrogen atom which absorbs the same number of XUV photons as the hydrogen molecule, we can identify the minimums resulting from the  destructive interference of the electron wave packets emitted from the hydrogen molecule, where these minimums are pointed out by arrows in Fig.~5. The corresponding ejection angle $\theta_f$ can be used in the transformation of Eq. (\ref{eq8}) to determine the nuclei distance of the diatomic molecule, i.e.,
\begin{equation}\label{eq9}
R_0=\sqrt{\frac{[(1+2n)\pi]^2}{2E_f\cos^2(\theta_f-\theta_m)}}.
\end{equation}
Additionally, comparing Fig. 5 with Fig. 1, one may find that the IR laser field can broaden the kinetic energy distribution of the ionized electron and thus a continuous angle-resolved ionization spectrum may be obtained in IR+XUV two-color laser fields.

\begin{center}
\includegraphics{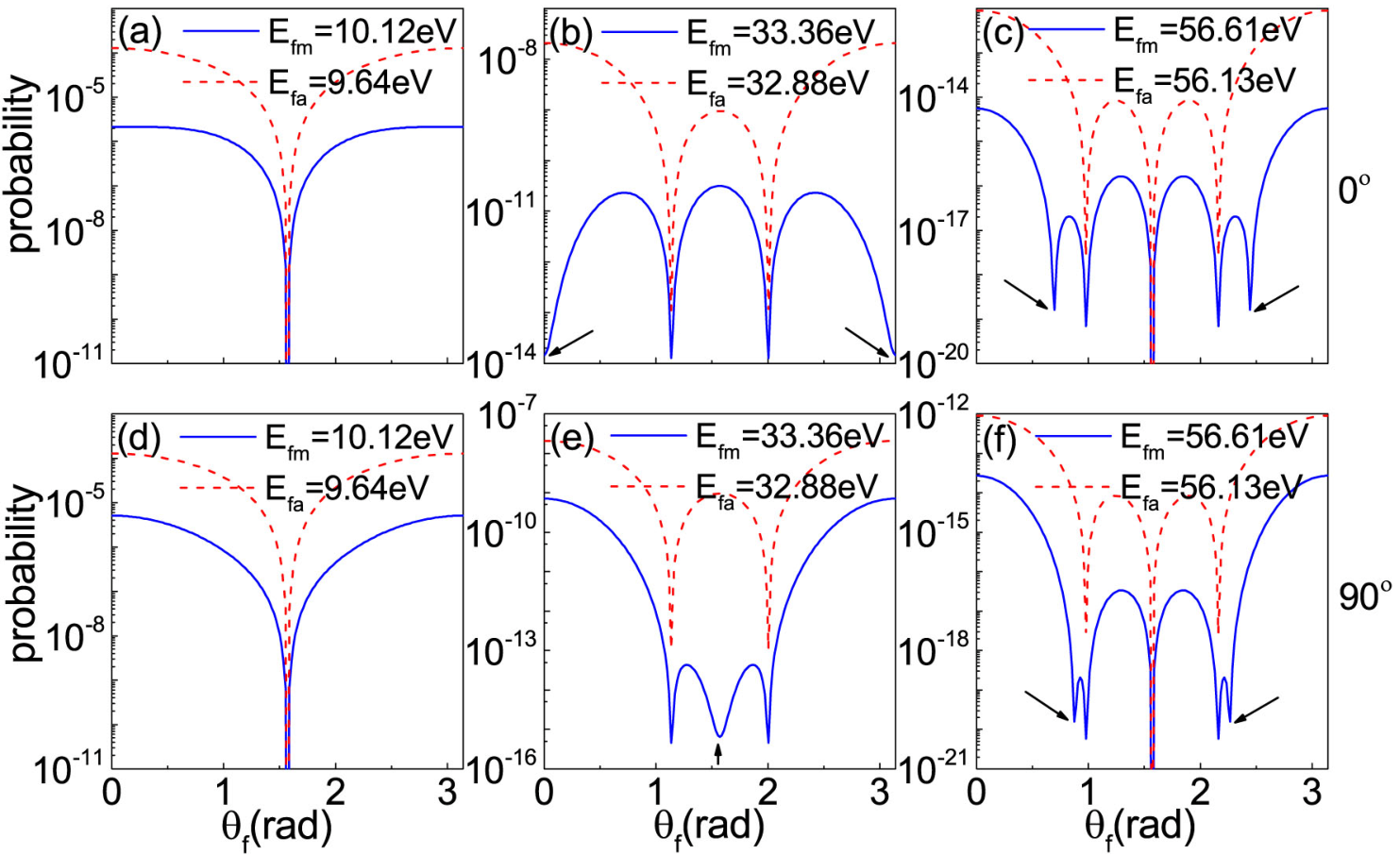}\\[5pt]  
\parbox[c]{15.0cm}{\footnotesize{\bf Fig.~5.} (color online) The angle-resolved probability  spectra for the direct ATI of a hydrogen molecule (dashed line in each panel) and a hydrogen atom (solid line in each panel) in a XUV laser filed with the intensity of ${\rm 3.6\times10^{13}~W/cm}^2$. $E_{fm}$ denotes the energy of the electron which is ionized from the molecular hydrogen and $E_{fa}$ denotes the energy of the electron which is ionized from the atomic hydrogen. For the three columns from left to right,  the electron absorbs one, two and three XUV photons, respectively. The molecular orientation angle $\theta_m$ is $0^{\circ}$ for the top panels and $90^{\circ}$ for the bottom panels. In logarithmic scale.}
\end{center}

We further calculate the momentum spectra for the direct ATI process of ${\rm H}_2$ in an XUV laser field with an intensity of ${\rm 3.6 \times 10^{14}~W/cm}^2$, which are plotted in Fig. 6. Here, the azimuthal angle of the final momentum of the electron $\varphi_f$ is variable, thus the momentum spectra exhibit a continuous distribution. We can find that there are two complete destructive interference fringes (CDIF) in the momentum spectra for each alignment angle of the molecular axis. For alignment angle $\theta_m$=~0$^\circ$, the CDIF are perpendicular to $p_z$ axis (see Figs. 6(a) and (c)), while for $\theta_m$=~90$^\circ$, the CDIF are parallel to $p_z$ axis (see Figs. 6(b) and (d)). Because the destructive condition $\cos(p_f\cos\theta R_0/2)=\cos[\frac{R_0}{2}(p_x\sin\theta_m+p_z\cos\theta_m)]=0$ determines the position of the CDIF at $p_z=\frac{(2n+1)\pi}{R_0}$ when $\theta_m$=~0$^\circ$, where $n=0, \pm1, \pm2, ...$, and at $p_x=\frac{(2n+1)\pi}{R_0}$ when $\theta_m$=~90$^\circ$, where $n=0, \pm1, \pm2, ...$. Hence, the molecular internuclear distance can be obtained by $R_0=\frac{2\pi}{\Delta p_j}$ $(j=x,z)$ once the momentum difference $\Delta p_z$ or $\Delta p_x$ is calculated from two adjacent CDIF in the momentum spectra.

\begin{center}
\includegraphics{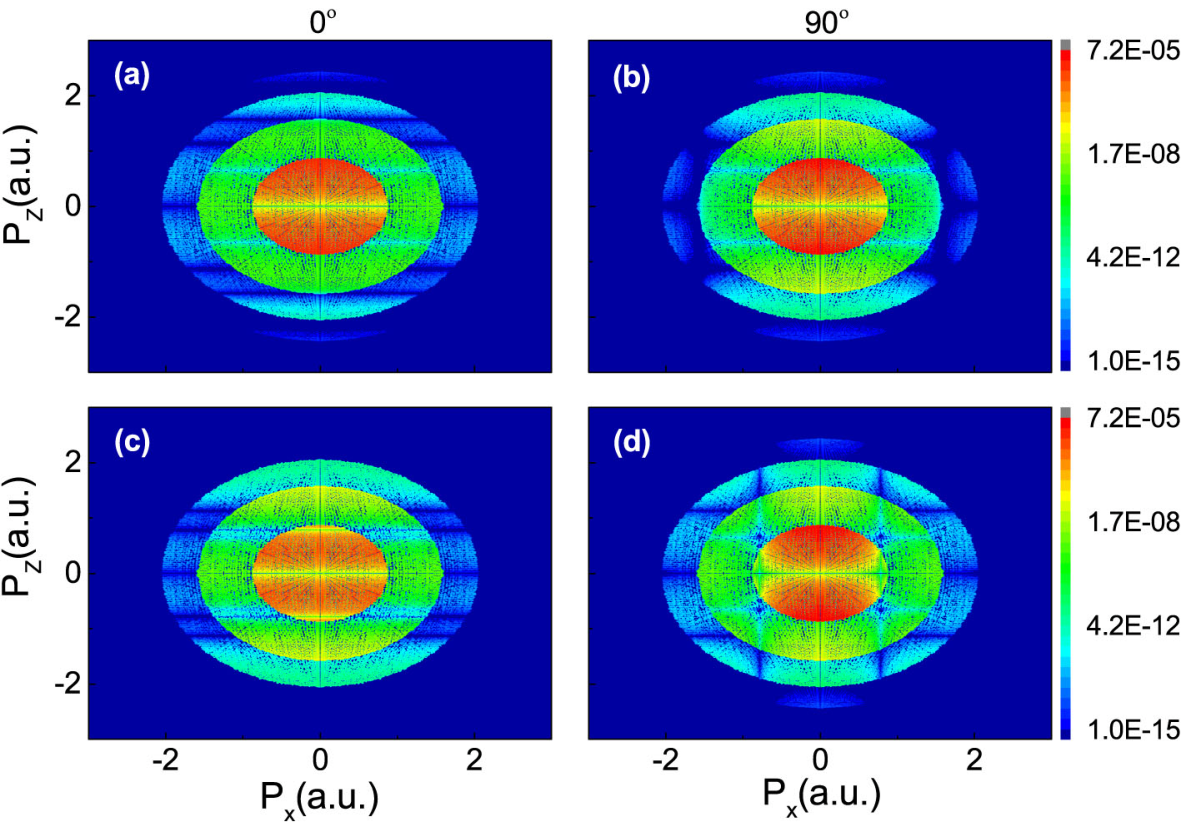}\\[5pt]  
\parbox[c]{15.0cm}{\footnotesize{\bf Fig.~6.} (color online) The ATI momentum spectra of a hydrogen molecule in a XUV laser field with an intensity of ${\rm 3.6\times10^{14}~W/cm}^2$. The upper panels are for the internuclear distance  $R_0=2.0~a.u.$ and the lower panels are for $R_0=4.0~a.u.$ The molecular axis is parallel (the left) or perpendicular (the right) to the polarization directions of the two laser fields. In logarithmic scale.}
\end{center}

In order to illustrate the applicability of the angle-resolved spectra of the direct ATI in IR+XUV laser fields to the exploration of the molecular structure, we  calculated the angular distribution of the direct ATI for ${\rm O}_2$ in IR+XUV laser fields with $\omega_2=50\omega_1$ and $\omega_1=1.55$~eV for the same intensities as in Fig. 1, as displayed in Fig. 7. The internuclear distance of ${\rm O}_2$ is 2.282 a.u. and the ionization potential is 12.07 eV. The wavefunction of ${\rm O}_2$ in momentum representation is formulated by Eq. (\ref{eq7}). By taking $\sin(p_f\cos\theta R_0/2)=0$, we have $p_f\cos\theta R_0/2=n\pi, n=0, \pm1, \pm2, ...$. When $n=0$, the equation  $\cos\theta=0$ yields a destructive stripe in the spectra at electronic emission angle  $\theta_f=\theta_m+\frac{\pi}{2}$, while when  $n=\pm1, \pm2, ...$, the predictive curve about the DIF can be expressed by the equation
\begin{equation}\label{eq10}
E_f=\frac{p_f^2}{2}=\frac{2n^2\pi^2}{R_0^2\cos^2(\theta_f-\theta_m)}.
\end{equation}
Meanwhile, by taking $\sqrt{1-\cos^2\theta}=0$, there will  also be some destructive stripes in the spectra at electronic emission angle $\theta_f=\theta_m~(\theta_m\geq0)$ or $\theta_f=\pi~(\theta_m=0)$. The black short-dashed lines in the spectra are the $E_f-\theta_f$ curves for $n=1$, and the blue dashed lines originate from $\cos\theta=0$ or $\cos^2\theta=1$. Both kinds of dashed lines fit well with the destructive fringes in the spectra. Additionally, in order to explore the source of the butterfly patterns in Fig. 7, we analyze the generalized Bessel function in Eq. (\ref{eq5}). We found that the term $J_{-q_3}(\zeta_3)$ is negligible under the laser conditions in this work, such that the generalized Bessel function can be written as
\begin{equation}\label{eq11}
\begin{array}{l}
\mathcal{J}_{q_1 q_2}(\zeta_f)\approx\sum\limits_{ q_4 q_5 q_6} J_{-q_1+q_5+q_6}(\zeta_1)J_{-q_2+2q_4+q_5-q_6}(\zeta_2)\\
~~~~~~~~~~~~~~~~~~~~\times J_{-q_4}(\zeta_4)J_{-q_5}(\zeta_5)J_{-q_6}(\zeta_6).
 \end{array}
\end{equation}
The red dash-dotted lines coinciding with the butterfly wing-shaped interference fringes are acquired from the minimum of $\mid\mathcal{J}_{q_1 q_2}(\zeta_f)\mid^2$, indicating that the butterfly wing-shaped interference fringes are attributed to the interaction between the laser fields and the ionized electron, irrelevant to the geometrical structure of molecules.

\begin{center}
\includegraphics{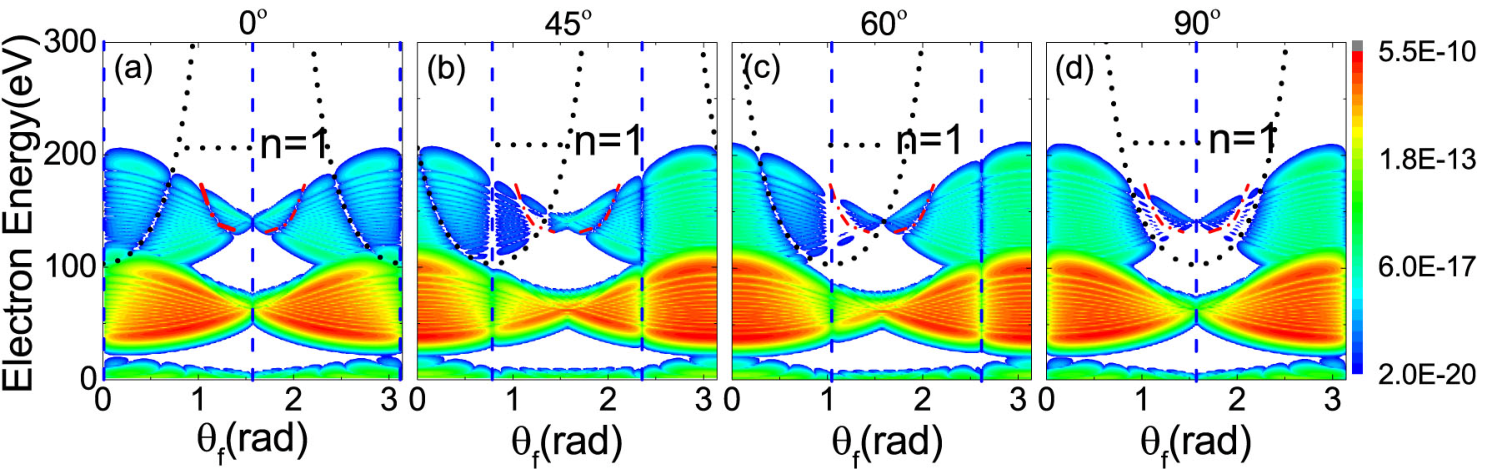}\\[5pt]  
\parbox[c]{15.0cm}{\footnotesize{\bf Fig.~7.} (color online) The angle-resolved direct ATI spectra of an oxygen molecule with an internuclear separation of 2.282~a.u. The photon energy of the XUV laser  $\omega_2=50\omega_1$, where $\omega_1=1.55$~eV. The laser intensities are $I_2=I_1={\rm 3.6\times10^{13}~W/cm}^2$. In logarithmic scale.}
\end{center}

\section{Conclusion}

By applying the frequency-domain theory based on the nonperturbative quantum electrodynamics, we have studied the direct ATI of  diatomic molecules in linearly polarized IR+XUV laser fields. The destructive interference fringes resulting from the coherent emission of the ionized electron are perfectly reproduced at any orientation angle by a simple predictive formula, which also predicts the minimal energy value for the emergence of the DIF in the spectra. The comparison between the direct ATI spectra of  a monochromatic laser field and two-color IR+XUV laser fields shows that the XUV laser field can not only raise the energy of the ionized electron, but also increase the ionization probability of the photoelectron in the high-energy region, while the IR laser field can broaden the kinetic energy distribution of the ionized electron, making the DIF clearly displayed in the direct ATI spectra. The study on the direct ATI of the oxygen molecule further demonstrates that the angle-resolved spectra of the direct ATI in IR+XUV laser fields may be extensively used in the investigation of molecular structures.

\addcontentsline{toc}{chapter}{Appendix A: Appendix section heading}
\section*{Appendix}
Under the calculated conditions, the width of each energy plateau is determined by the generalized Bessel function $J_{-q_1}(\zeta_{1},\zeta_{3})$, which can  be expressed in the form
\begin{equation}\label{eq13}
\begin{array}{l}
J_{-q_1}(\zeta_{1},\zeta_{3})=\frac{1}{T}\int_{-T/2} ^{T/2} dt\exp[if(t)]
  \end{array}
\end{equation}
where $f(t)=\zeta_{1}\sin(\omega_1t)+\zeta_{3}\sin(2\omega_1t)+q_1\omega_1t$ and $T=2\pi/\omega_1$. On the other hand, the classical action of an electron is
\begin{equation}\label{eq14}
\begin{array}{l}
S_{cl}(t,\textbf{p}_f)=\frac{1}{2}\int_0 ^t dt^{'} [\textbf{p}_f+\textbf{A}_{cl}(t^{'})]^2,
 \end{array}
\end{equation}
where $\textbf{A}_{c1}(t)=\hat{\epsilon}E_0/\omega_1\cos(\omega_1t)$ is the vector potential of the XUV laser field~\cite{Zhang2013}. Comparing Eq.(\ref{eq13}) with Eq. (\ref{eq14}), $J_{-q_1}(\zeta_{1},\zeta_{3})$ can be rewritten as
\begin{equation}\label{eq15}
\begin{array}{l}
J_{-q_1}(\zeta_{1},\zeta_{3})=\frac{1}{T}\int_{-T/2} ^{T/2} dt\exp\{i[S_{cl}(t,\textbf{p}_f)-(q_2\omega_2-I_p)t]\}

 \end{array}
\end{equation}
In the saddle-point approximation, the saddle-point $t_0$ satisfies $f'(t_0)=0$, which leading to the energy conservation relation in the ionization process
\begin{equation}\label{eq16}
\frac{[\textbf{p}_f+\textbf{A}_{c1}(t)]^2}{2}=q_2\omega_2-I_p.
\end{equation}
This equation can predict the begining and cutoff position of the ATI spectrum for each $q_2$. When $\theta_f<\frac{\pi}{2}$, the maximum energy value is
\begin{equation}\label{eq17}
\begin{array}{l}
E_{fmax}=(2\sqrt{\widetilde{U}_{p_1}}\cos(\theta_f)+\sqrt{4\widetilde{U}_{p_1}\cos^2\theta_f+2(q_2\omega_2-I_p-2\widetilde{U}_{p_1})}~)^2/2
 \end{array}
\end{equation}
under the condition of $\cos\omega_1t_0=1$, and the minimum energy value is
\begin{equation}\label{eq18}
\begin{array}{l}
E_{fmin}=(-2\sqrt{\widetilde{U}_{p_1}}\cos(\theta_f)+\sqrt{4\widetilde{U}_{p_1}\cos^2\theta_f+2(q_2\omega_2-I_p-2\widetilde{U}_{p_1})}~)^2/2
 \end{array}
\end{equation}
under the condition of $\cos\omega_1t_0=-1$. When $\theta_f>\frac{\pi}{2}$, the maximum energy value is
\begin{equation}\label{eq19}
\begin{array}{l}
E_{fmax}=(-2\sqrt{\widetilde{U}_{p_1}}\cos(\theta_f)+\sqrt{4\widetilde{U}_{p_1}\cos^2\theta_f+2(q_2\omega_2-I_p-2\widetilde{U}_{p_1})}~)^2/2
 \end{array}
\end{equation}
where $\cos\omega_1t_0=-1$, and the minimum energy value is
\begin{equation}\label{eq20}
\begin{array}{l}
E_{fmin}=(2\sqrt{\widetilde{U}_{p_1}}\cos(\theta_f)+\sqrt{4\widetilde{U}_{p_1}\cos^2\theta_f+2(q_2\omega_2-I_p-2\widetilde{U}_{p_1})}~)^2/2
 \end{array}
\end{equation}
where $\cos\omega_1t_0=1$.

\addcontentsline{toc}{chapter}{Acknowledgment}
\section*{Acknowledgment}
 We thank all the members of SFAMP club for helpful discussions.




\begin{thebibliography}{99}\footnotesize
\itemsep=-3pt plus.2pt minus.2pt   

\bibitem{chu} Heslar J and Chu S I \href{https://journals.aps.org/pra/abstract/10.1103/PhysRevA.95.043414}{2017 \emph{Phys. Rev. A} \textbf{95} 043414}

\bibitem{gao2000} Gao L, Li X and Fu P \href{https://journals.aps.org/pra/abstract/10.1103/PhysRevA.61.063407}{2000 \emph{Phys. Rev. A} \textbf{61} 063407}

\bibitem{DBM2018} Busulad\v{z}i\'{c} M, \v{C}erki\'{c} A, Gazibegovi\'{c}-Busulad\v{z}i\'{c} A,  Hasovi\'{c} E and  Milo\v{s}evi\'{c} D B \href{https://journals.aps.org/pra/abstract/10.1103/PhysRevA.98.013413}{2018 \emph{Phys. Rev. A} \textbf{98} 013413}

\bibitem{Dino2018}  Habibovi\'{c} D,  \v{C}erki\'{c} A,  Busulad\v{z}i\'{c} M,  Gazibegovi\'{c}-Busulad\v{z}i\'{c} A,  Od\v{z}ak S,  Hasovi\'{c} E,
 Milo\v{s}evi\'{c} D B \href{https://doi.org/10.1007/s11082-018-1480-6}{2018 \emph{Opt Quant Electron}  \textbf{50} 214}

\bibitem{Wang2007}  Wang B, Gao L, Li X, Guo D S and  Fu P \href{https://journals.aps.org/pra/abstract/10.1103/PhysRevA.75.063419}{2007 \emph{Phys. Rev. A} \textbf{75} 063419}

\bibitem{chun2009} Guo Y C, Fu P M and Wang B B \href{http://iopscience.iop.org/article/10.1088/0256-307X/26/3/034204/meta}{2009 \emph{Chin. Phys. Lett.} \textbf{26} 034204}

\bibitem{Guo2009}  Guo Y, Fu P, Yan Z C, Gong J and  Wang B \href{https://journals.aps.org/pra/abstract/10.1103/PhysRevA.80.063408}{2009 \emph{Phys. Rev. A} \textbf{80} 063408}

\bibitem{becker1} Busulad\v{z}i\'{c} M,  Gazibegovi\'{c}-Busulad\v{z}i\'{c} A,  Milo\v{s}evi\'{c} D B and Becker W \href{https://journals.aps.org/prl/abstract/10.1103/PhysRevLett.100.203003}{2008 \emph{Phys. Rev. Lett.} \textbf{100} 203003}

\bibitem{Hetzheim2007} Hetzheim H, Figueira de Morisson Faria C and  Becker W \href{https://journals.aps.org/pra/abstract/10.1103/PhysRevA.76.023418}{2007 \emph{Phys. Rev. A} \textbf{76} 023418}

\bibitem{kang2010}  Kang H,  Quan W, Wang Y,  Lin Z,  Wu M,  Liu H,  Liu X,  Wang B B,  Liu H J,  Gu Y Q,  Jia X Y,
 Liu J,  Chen J and  Cheng Y \href{https://journals.aps.org/prl/abstract/10.1103/PhysRevLett.104.203001}{2010 \emph{Phys. Rev. Lett.} \textbf{104} 203001}

\bibitem{becker2} Okunishi M,  Itaya R,  Shimada K,  Pr\"{u}mper G,  Ueda K,  Busulad\v{z}i\'{c} M,  Gazibegovi\'{c}-Busulad\v{z}i\'{c} A,
 Milo\v{s}evi\'{c} D B and  Becker W \href{https://journals.aps.org/prl/abstract/10.1103/PhysRevLett.103.043001}{2009 \emph{Phys. Rev. Lett.} \textbf{103} 043001}

\bibitem{Wang2010}  Wang B,  Guo Y,  Zhang B,  Zhao Z,  Yan Z C and  Fu P \href{https://journals.aps.org/pra/abstract/10.1103/PhysRevA.82.043402}{2010 \emph{Phys. Rev. A} \textbf{82} 043402}

\bibitem{ingo2015}  Petersen I,  Henkel J and  Lein M \href{https://journals.aps.org/prl/abstract/10.1103/PhysRevLett.114.103004}{2015 \emph{Phys. Rev. Lett.} \textbf{114} 103004}

\bibitem{Odzak2014}  Od\v{z}ak S,  \v{C}erki\'{c} A,  Busulad\v{z}i\'{c} M,  Hasovi\'{c} E,  Gazibegovi\'{c}-Busulad\v{z}i\'{c} A and  Milo\v{s}evi\'{c} D B \href{http://iopscience.iop.org/article/10.1088/0031-8949/2014/T162/014012/meta}{2014 \emph{Phys.~Scr.} \textbf{162} 014012}

\bibitem{toru2008}  Morishita T,  Le A T,  Chen Z and  Lin C D \href{https://journals.aps.org/prl/abstract/10.1103/PhysRevLett.100.013903}{2008 \emph{Phys. Rev. Lett.} \textbf{100} 013903}

\bibitem{odzak2009}  Od\v{z}ak S and  Milo\v{s}evi\'{c} D B \href{http://iopscience.iop.org/article/10.1088/0953-4075/42/7/071001/meta}{2009 \emph{J.~Phys.~B} \textbf{42} 071001}

\bibitem{Augstein2011}  Augstein B B and   de Morisson Faria C F \href{http://iopscience.iop.org/article/10.1088/0953-4075/44/5/055601/meta}{2011 \emph{J.~Phys.~B} \textbf{44} 055601}

\bibitem{Su2018} Su N,  Yu S, Li W, Yang S and  Chen Y  \href{http://cpb.iphy.ac.cn/article/2018/1940/cpb_27_5_054213.html}{2018 \emph{Chin. Phys. B} \textbf{27} 054213}

\bibitem{Torres2010} Torres R,  Siegel T,  Brugnera L,  Procino I,  Underwood J G,  Altucci C,  Velotta R,  Springate E,  Froud C,
 Turcu I C E,  Patchkovskii S,  Ivanov M Y,  Smirnova O, and  Marangos J P \href{https://journals.aps.org/pra/abstract/10.1103/PhysRevA.81.051802}{2010 \emph{Phys. Rev. A} \textbf{81} 051802}

\bibitem{EBERLY1991}  Eberly J H,  Javanainen J,  Rz\c{a}\.{z}ewski K \href{https://doi.org/10.1016/0370-1573(91)90131-5}{1991 \emph{Phys. Rep.} \textbf{204} 331¡ª383}

\bibitem{BECKER2008}  Becker W,  Grasbon F,  Kopold R,  Milo\v{s}evi\'{c} D B,  Paulus G G and  Walther H \href{https://doi.org/10.1016/S1049-250X(02)80006-4}{2008 \emph{Adv. Atom. Mol. Opt. Phys.} \textbf{48} 35-98}

\bibitem{he2015}  He P L and  He F \href{http://iopscience.iop.org/article/10.1088/0031-8949/90/4/045402/meta}{2015 \emph{Phys.~Scr.} \textbf{90}  045402}

\bibitem{kui2017}  Zhang K,  Liu M,  Wang B B,  Guo Y C,  Yan Z C,  Chen J,  Liu X J \href{http://iopscience.iop.org/article/10.1088/0256-307X/34/11/113201/meta}{2017 \emph{Chin. Phys. Lett.} \textbf{34} 113201}

\bibitem{Richard2008}  Ta\"{i}eb R,  Maquet A and  Meyer M \href{http://iopscience.iop.org/article/10.1088/1742-6596/141/1/012017/meta}{2008 \emph{J. Phys.: Conf. Ser.} \textbf{141} 012017}

\bibitem{Liu2015}  Liu M,  Guo Y C and  Wang B  \href{http://iopscience.iop.org/article/10.1088/1674-1056/24/7/073201/meta}{2015 \emph{Chin. Phys. B} \textbf{24} 073201}

\bibitem{Zhang2013} Zhang K,  Chen J,  Hao X L,  Fu P, Yan Z C and  Wang B \href{https://journals.aps.org/pra/abstract/10.1103/PhysRevA.88.043435}{2013 \emph{Phys. Rev. A} \textbf{88} 043435}

\bibitem{Guo1992}  Guo D S and  Drake G W F \href{http://iopscience.iop.org/article/10.1088/0305-4470/25/20/018/meta}{1992 \emph{J. Phys. A} \textbf{25} 5377}



\end{thebibliography}
\end{document}